\documentclass[conference]{IEEEtran}
\ifCLASSINFOpdf
\else
\fi
\usepackage{amsmath,bm,graphicx}
\usepackage{algorithm,algpseudocode}
\begin{document}
%
\title{A Robust $\ell_1$ Penalized DOA Estimator}

\author{ Ashkan Panahi and Mats Viberg\\
Signal processing group, Signals and Systems department\\
Chalmers university of technology, Gothenburg, Sweden
}


%


\maketitle

\begin{abstract}
The SPS-LASSO has recently been introduced as a solution to the problem of regularization parameter selection in the complex-valued LASSO problem. Still, the dependence on the grid size and the polynomial time of performing convex optimization technique in each iteration, in addition to the deficiencies in the low noise regime, confines its performance for Direction of Arrival (DOA) estimation. This work presents  methods to apply LASSO without grid size limitation and with less complexity. As we show by simulations, the proposed methods loose a negligible performance compared to the Maximum Likelihood (ML) estimator, which needs a combinatorial search We also show by simulations that compared to practical implementations of ML, the proposed techniques are less sensitive to the source power difference. 
	
\end{abstract}


%
\IEEEpeerreviewmaketitle

\newcommand{\argmin}{\operatornamewithlimits{argmin}}
\newcommand{\argmax}{\operatornamewithlimits{argmax}}
\newcommand{\x}{\mathbf{x}}
\newcommand{\nhat}{\hat{\mathbf{n}}}
\newcommand{\as}{\mathbf{a}}
\newcommand{\A}{\mathbf{A}}
\newcommand{\btheta}{\bm{\theta}}
\newcommand{\s}{\mathbf{s}}
\newcommand{\n}{\mathbf{n}}
\newcommand{\shat}{\hat{\mathbf{s}}}
\newcommand{\bthetahat}{\hat{\bm{\theta}}}
\newcommand{\re}{\text{Re}}
\newcommand{\ds}{\mathbf{d}}
\newcommand{\betta}{\bm{\eta}}
\newcommand{\J}{\mathbf{J}}
\newcommand{\js}{\mathbf{j}}
\newcommand{\D}{\mathbf{D}}
\newcommand{\Sd}{\mathbf{S}}
\newcommand{\bGamma}{\bm{\Gamma}}
\newcommand{\zero}{\mathbf{0}}
\section{Introduction}

The Least Absolute Shrinkage and Selection Operator (LASSO) is a method of choosing a small set of bases among a large collection, best representing a set of data linearly. It is based on $\ell_1$ regularization of an Ordinary Least Square (OLS), which always gives a sparse solution based on the conic nature of the cost function \cite{tibshirani}. The consistency of such a technique, under the assumption of the "best collection", is well discussed in an asymptotic case that both the data dimension and the size of collection increase to infinity \cite{Donoho_CS}. 

A much different attempt has been made by applying LASSO into a finite dimensional data set with an asymptotically large set of basis dictated by a physical model. A well known example is in \cite{Malioutov}, where the LASSO technique is applied to the problem of estimating Direction Of Arrivals (DOAs) which can be expressed and solved in a linear regression fashion by discretizing the DOA parameter space. This method gained more attention since the solution could be found robustly, independent of the choice of the initial values in the numerical optimization technique due to the convex nature of the cost function.

Unlike the statistical regression application of LASSO, the choice of the Regularization Parameter (RP) is critical in the current application. This parameter implements the trade off between model precision and model order. There are numerous ways to estimate the RP when the true number of sources is unknown \cite{Panahi1,BLASSO}. However, the estimation quality is always improved by choosing a smaller RP value with the same model order. On the other hand, finding the smallest such parameter is not straightforward, since the implementation of LASSO by convex programming techniques \cite{cvx} is computationally costly, and can not be performed for a fine search over possible RP values. A stagewise solution is found in \cite{LARS} for the real regression case, by the observing that the homotopy path of LASSO solutions is piecewise linear \cite{Homotopy}. Later, in \cite{panahi_fast} the idea was generalized to the complex problems by introducing a different LASSO optimality condition which is called Singular Point Selection (SPS)-LASSO. Decreasingly in RP, the SPS-LASSO  follows the points in the homotopy path, in which a new regressor is born. These points are known as the candidate points.

The SPS-LASSO technique is not suitable for the DOA estimation problem, especially in low SNR cases, due to the fact that the solution support moves continuously between two singular points. Furthermore, the discretized nature of the problem introduces an additional quantization error. Accordingly, in this work, we introduce an alternative recursive solution, by keeping only the estimate support at each iteration and modifying the optimality conditions to avoid a discretized space. The resulting algorithm does not have the initialization problems of NonLinear Least Squares (NLLS) and is faster than convex programming techniques. We further simplify the algorithm to get a faster solution. We show by simulations that the faster version is also convergent to the optimal point with very high probability. The results show that the proposed optimal point is very close to the ML global minimum point, so that we loose a negligible performance compared to ML. Note that ML is very costly to be implemented when the model order grows. We further compare LASSO to other approximations of ML such as Space alternating Generalized EM (SAGE \cite{SAGE}).

\section{Problem statement}
This work concerns the problem of DOA estimation with an array of $m$ sensors, receiving one snapshot of narrowband signals from $n$ far sources. Due to the far-field model of the transmitting wave, the received signal pattern is mostly defined by phase shifts at each sensor so that the received signal vector $\x$ can be written as (\cite{Viberg_2decades})
\begin{equation}
\x=\sum_{i=1}^n{\as(\theta_i)s_i}+\n=\A(\btheta)\s+\n,
\end{equation} 
where $\s=[s_1\ s_2\ \ldots s_n]^T$, $\btheta=[\theta_1\ \theta_2\ \ldots\theta_n]$, and $\n$ are the source waveform, DOA, and measurement noise vectors respectively. Furthermore, $\as(\theta)$ is the steering vector, which represents the phase shift operations at different sensors corresponding to the DOA $\theta$. The problem is to estimate the DOA vector $\btheta$ given the measurement vector $\x$ assuming an uncorrelated, circularly symmetric, centered Gaussian noise vector $\n$.

\subsection{Conventional Solutions}
Under the above statistical assumptions of the noise vector, the deterministic ML estimator is given by (\cite{Viberg_2decades})
\begin{equation}\label{eq:NLLS}
(\bthetahat,\shat)=\argmin_{(\btheta,\s)}\|\x-\A(\btheta)\s\|_2^2,
\end{equation}
where $\btheta\in[0\ \pi]^n$. The solution of \ref{eq:NLLS} could be found by noting its corresponding Karush-Kuhn-Tucker (KKT) conditions (\cite{Bazaraa})
\begin{equation}\label{eq:KKTml}
\forall\theta_i\in\btheta:\left\{\begin{array}{l}
\as^H(\theta_i)\nhat=0\\
\Re(s_i^*\ds^H(\theta_i)\nhat)=0.
\end{array}\right.
\end{equation}
where $\nhat=\x-\A(\btheta)\s$ and $\mathbf{d}(\theta)=\frac{\partial\as(\theta)}{\partial\theta}$. The system (\ref{eq:KKTml}) does not have a unique solution although the ML global optimum is unique. However, starting from a sufficiently close point, and following a Newton recursive algorithm, the solution may converge to the global optimum.  We  introduce Algorithm \ref{alg:ML} as a typical such solution in which we write the complex waveforms in polar coordinates as $s_i=r_ie^{j\alpha_i}$ and $\J_\text{ML}$ is given in (\ref{eq:Jml}) where $\bm{\Delta}_1$ and $\bm{\Delta}_2$ are diagonal matrices who's diagonal elements are given by $\ds^H(\theta_i)\nhat$ and $\mathbf{c}^H(\theta_i)\nhat$ with $\mathbf{c}(\theta)=\frac{\text{d}\ds(\theta)}{\text{d}\theta}$, respectively. Furtheremore, $\D(\btheta)=[\ds(\theta_1)\ \ds(\theta_2)\ldots\ds(\theta_n)]$ and $\Sd$ and $\bGamma$ are diagonal matrices who's diagonal elements are $s_i$ and $e^{j\alpha_i}$ respectively. More details of such methods could be found in \cite{Newtonian}. 
\begin{algorithm}
\caption{ML estimator Newton solver}
\label{alg:ML}
\begin{algorithmic}
\State $\btheta\gets\btheta_0$ and $\s\gets\s_0$.
\While {not converging}
\State $\nhat\gets\x-\A(\btheta)\s$
\State $\betta_\text{ML}\gets\left[\Re(\A^H(\btheta)\nhat)^T\ \Im(\A^H(\btheta)\nhat)^T\  \Re(\D^H(\btheta)\nhat)^T\right]^T$
\State $\betta_0\gets\J_\text{ML}^{-1}\betta_\text{ML}$
\For{i=1:n} 
\State $\theta_i\gets\theta_i+\eta_{0,i}$ 
\State $r_i\gets r_i+\eta_{0,n+i}$
\State $\alpha_i\gets\alpha_i+\eta_{0,2n+i}$
\EndFor
\EndWhile 
\end{algorithmic}
\end{algorithm}

The convergence could be checked simply by thresholding the difference in the sequence of estimates or the sequence of the cost function. Although the Algorithm \ref{alg:ML} is naive in practice, due to the inversion instability of the Jacobian matrix $\J$, we introduce it as a base of development toward our proposed method. 
\begin{figure*}
\centering
\begin{eqnarray}\label{eq:Jml}
&\J_\text{ML}=\left[\begin{array}{ccc}
\Re(\bm{\Delta_1}-\A^H(\btheta)\D(\btheta)\Sd)&
-\Re(\A^H(\btheta)\A(\btheta)\bGamma)& \Im(\A^H(\btheta)\A(\btheta)\Sd)\\
\Im(\bm{\Delta_1}-\A^H(\btheta)\D(\btheta)\Sd)&
-\Im(\A^H(\btheta)\A(\btheta)\bGamma)& -\Re(\A^H(\btheta)\A(\btheta)\Sd)\\
\Re(\bm{\Delta_2}-\D^H(\btheta)\D(\btheta)\Sd)&
-\Im(\D^H(\btheta)\A(\btheta)\bGamma)& -\Re(\D^H(\btheta)\A(\btheta)\Sd)\\
\end{array}\right]\nonumber\\
&\js=\left[\begin{array}{ccc}
-\D(\btheta)\Sd&
-\A(\btheta)\bGamma&
 -j\A(\btheta)\Sd
\end{array}\right]
\end{eqnarray}
\end{figure*}
As we have already explained, the direct ML realization is costly when the desired model order is high. Furtheremore, such method is not stable due to the .  

There has been a variety of approximate, low-cost solutions to ML such as Expectation-Maximization (EM) algorithms and Space Alternating Generalized EM (SAGE \cite{SAGE}) as well as RELAX \cite{RELAX} all more or less dependent on the choice of the initial values. As another alternative, the possibility of estimating DOAs using $\ell_1$ penalized Ordinary Least Squares (OLS) has been known and discussed for almost a decade (\cite{Malioutov}). This method solves the problem of local minima by minimizing an approximate convex cost function which is independent of the initial values. Assume a discretization $\btheta^g=\{\theta^g_1,\theta^g_2,\ldots,\theta^g_N\}$ of $[0\ \pi]$. According to \cite{Malioutov}, the solution of (\ref{eq:NLLS}) could be approximated by first solving
\begin{equation}
\shat^g=\argmin_{s^g}\frac{1}{2}\|\x-\A^g\s^g\|_2^2+\lambda\|\s^g\|_1,
\end{equation}
where $\A^g=\A(\btheta^g)$ and $\lambda$ is a regularization parameter \cite{Panahi1}, and next introducing $\bthetahat$ as the elements in $\btheta^g$ with nonzero corresponding $\shat^g$ elements.

The parameter $\lambda$ sets the compromise between the level of sparseness and model fit, and is usually hard to determine analytically. On the other hand, realizing LASSO in (\ref{eq:NLLS}) for a big set of $\lambda$ values is infeasible. In \cite{panahi_fast}, an alternative solution is introduced by the SPS-LASSO stagewise algorithm, which only needs LASSO realizations at a recursively determined finite set of candidate $\lambda$ values. This could be done by observing the following optimality conditions of LASSO \cite{panahi_fast}
\begin{eqnarray}
\forall\theta^g_i\in\btheta^g :\left\{\begin{array}{l}
|\as^H(\theta^g_i)\nhat|\leq\lambda\\
\theta^g_i\in I\Rightarrow\as^H(\theta)\nhat=\lambda\frac{\hat{s_i}}{|\hat{s_i}|}
\end{array}\right.,
\end{eqnarray}
where $I=\{\theta_i^g|\hat{s}_i^g\neq 0\}$ and $\nhat=\x-\A^g\s$. Let us define the following function:
\subsubsection{Marginalized Source Attractor (G)} Given the quadruple $(I,r,\alpha,\lambda)$ and a point $\theta$, the function $G(I,r,\alpha,\phi;\theta)$ gives the convergence point $(r^\prime,\alpha^\prime,\lambda^\prime)$ of the Newton iterative solution of the following system of equations starting from  $(r,\alpha,\lambda)$ with $s_i=r_ie^{j\alpha_i}$.
\begin{eqnarray}\label{eq:Gfunction}
\as^H(\theta_i)\nhat^\prime=\lambda^\prime e^{j\alpha_i^\prime}\nonumber\\
|\as^H(\theta_i)\nhat^\prime|=\lambda^\prime.
\end{eqnarray}
where $\nhat^\prime=\x-\A(I)\s^\prime$. The G function could also be computed faster using an iterative algorithm as in \cite{panahi_fast}, but we will not express it here.  The algorithm could be written as Algorithm \ref{alg:SPS-LASSO} where $G_\lambda$ denotes the $\lambda$ element of the function $G$. Note that it is assumed that between each two candidate points in the solution path of LASSO (as a function of $\lambda$) the DOA estimate is constant.
\begin{algorithm}
\caption{SPS LASSO}
\label{alg:SPS-LASSO}
\begin{algorithmic}
\State $\theta_0\gets\argmax\limits_{\theta^g\in\btheta^g}|\as^H(\theta^g)\x|$
\State $I\gets\{\theta_0\}$ and $\lambda\gets|\as^H(\theta_0)\x|$ 
\State $r\gets 0$ and $e^{j\alpha}\gets\frac{\as^H(\theta_0)\x}{|\as^H(\theta_0)\x|}$
\State counter $\gets$ 0.
\While {$\text{counter} < n$}
\State $\theta_1\gets\argmax\limits_{\theta^g\in\btheta^g} G_\lambda(I,r,\alpha,\lambda;\theta^g)$ 

\State $I\gets I\cup{\theta_1}$ 
\State $(r,\alpha,\lambda)\gets G(I,r,\alpha,\lambda;\theta_1)$
\State $r_{\theta_1}\gets 0$ and $e^{j\alpha(\theta_1)}\gets\frac{\as^H(\theta_0)\x}{|\as^H(\theta_0)\x|}$
\State counter $\gets$ counter+1.
\EndWhile 
\end{algorithmic}
\end{algorithm}

 
The performance of SPS-LASSO is still limited by the size of the grid which may not be increased arbitrarily due to the complexity of convex optimization techniques \cite{cvx}. Furthermore, SPS-LASSO works incorrectly in very high SNR regimes, where the desired regularization parameter is extremely small and the approximately fixed DOAs assumption does not hold.

To overcome such difficulties, in this work we redefine the parameter space as the space of low dimensional non sparse vectors. However, inspired by the optimality conditions of LASSO, we introduce slightly modified feasible conditions which do not depend on any discretization. We demonstrate the details of two algorithms to find the unique optimal point of these new conditions. 

\section{Continuous LASSO}

Inspired by SPS-LASSO, we look for a finite subset $I=\{\theta_1,\theta_2,\ldots,\theta_r\}\subset[0\ \pi]$ and a complex function $s_i=s(\theta_i)$ on $I$ satisfying
\begin{eqnarray}\label{eq:CLASSO}
\forall\theta\in[0\ \pi]:\left\{\begin{array}{l}
|\as^H(\theta)\nhat|\leq\lambda\\
\theta=\theta_i\in I\Rightarrow\as^H(\theta)\nhat=\lambda\frac{{s_i}}{|{s_i}|}
\end{array}\right.,
\end{eqnarray}
where $\nhat=\x-\A(I)\s$. We also write $s=re^{j\alpha}$ where $(r,\alpha)$ are the polar coordinates of the function $s$. The existence and uniqueness of such set as well as consistency conditions of such method is proved but will not be presented in this work.
 
Finding such a set is guaranteed by the homotopy rule and the similarities to convex LASSO based estimation. Note that finding the solution for a fixed $\lambda$ by a gradient descent algorithm is not possible due to the infinite dimensional nature of the problem. Thus, we assume that $\lambda$ is variable. Note that for the values of $\lambda>\max\limits_\theta{|\as^H(\theta)\x|}$ the solution is given by $I=\emptyset$. Note that conditions (\ref{eq:CLASSO}) imply that each $\theta_i\in I$ is a global maximum point for the function $f(\theta)=|\as^H(\theta)\nhat|$. Thus, its derivative is zero which after some manipulations results in the condition
\begin{equation}\label{eq:mid}
\Re(s_i^*\ds^H(\theta_i)\nhat)=0.
\end{equation}
The equation (\ref{eq:mid}) and the second line of (\ref{eq:CLASSO}) might be compared to the ML optimality conditions in (\ref{eq:KKTml}). The main idea in this work is to relax the assumption that DOAs are fixed in the smooth homotopy pieces. Roughly speaking, this means that the marginalized source attractor $G$ should be modified to include DOA changes imposed by (\ref{eq:mid}) as well as source changes. Accordingly, we modify $G$ as follows:

\subsubsection{Local Marginalized Attractor ($F$)}: Given a quadruple $(I,r,\alpha,\lambda)$ and a point $\theta$, the function $F(I,r,\phi,\lambda;\theta)$ gives the convergence point quadruple $(I^\prime,r^\prime,\alpha^\prime,\lambda^\prime)$ starting from $(I,r,\alpha,\lambda)$ and following the Newton recursive solution of the system of equations
\begin{eqnarray}
&\as^H(\theta^\prime_i)\nhat^\prime=\lambda^\prime\frac{\s^\prime_i}{|\s^\prime_i|}
\nonumber\\
&\Re\left(\frac{{\s^\prime_i}^*}{|\s^\prime_i|}\mathbf{d}^H(\theta^\prime_i)\nhat^\prime\right)=0
\nonumber\\
&|\as^H(\theta)\nhat^\prime|=\lambda^\prime,
\end{eqnarray}
where $\nhat^\prime=\x-\A(I')\s^\prime$. Note that $F$ is obtained from $G$ by relaxing the DOA parameters and imposing (\ref{eq:mid}). The $G$ function could be found using a Newton algorithm similar to Algorithm \ref{alg:ML} by substituting $\J_\text{ML}$ and $\betta_\text{ML}$ with $\J_\text{LMA}$ and $\betta_\text{LMA}$ respectively where
\begin{equation}\label{eq:ettaLMA}
\betta_\text{LMA}=\left[\begin{array}{c}
\betta_\text{LEA}\\
|\as^H(\theta)\nhat|^2-\lambda^2
\end{array}\right],
\end{equation}
and
\begin{equation}\label{eq:JLMA}
\J_\text{LMA}=\left[\begin{array}{cc}
\J_\text{LEA}& \begin{array}{c}
\Re(e^{j\alpha_1})\\
\Re(e^{j\alpha_2})\\
\vdots\\
\Re(e^{j\alpha_n})\\
\Im(e^{j\alpha_1})\\
\Im(e^{j\alpha_2})\\
\vdots\\
\Im(e^{j\alpha_n})\\
\end{array}\\
\Re\left(\as^H(\theta)\js\nhat^H\as(\theta)\right) & -2\lambda
\end{array}\right],
\end{equation}
with $\js$ given in (\ref{eq:Jml}),
\begin{equation}\label{eq:ettaLEA}
\betta_\text{LEA}=\betta_\text{ML}-\lambda[\Re(e^{j\alpha})\ \Im(e^{j\alpha})\ 0]^T,
\end{equation}
and 
\begin{equation}\label{eq:JLEA}
\J_\text{LEA}=\J_\text{ML}+\left[\begin{array}{ccc}
\zero & \zero & \lambda\Im(\bGamma)\\
\zero & \zero & -\lambda\Re(\bGamma)\\
\zero & \zero & \zero
\end{array}\right].
\end{equation} 
The proposed C-LASSO algorithm is summarized as \ref{alg:C-LASSO}. below.
\begin{algorithm}
\caption{C-LASSO}
\label{alg:C-LASSO}
\begin{algorithmic}
\State $\theta_0\gets\argmax\limits_{\theta}|\as^H(\theta)\x|$
\State $I\gets\{\theta_0\}$ and $\lambda\gets|\as^H(\theta_0)\x|$ 
\State $r\gets 0$ and $e^{j\alpha}\gets\frac{\as^H(\theta_0)\x}{|\as^H(\theta_0)\x|}$
\State counter $\gets$ 0.
\While {$\text{counter} < n$}
\State $\theta_1\gets\argmax\limits_{\theta} F_\lambda(I,r,\alpha,\lambda;\theta)$ 

\State $(I, r,\alpha,\lambda)\gets F(I,r,\alpha,\lambda;\theta_1)$
\State $I\gets I\cup{\theta_1}$ 
\State $r_{\theta_1}\gets 0$ and $e^{j\alpha(\theta_1)}\gets\frac{\as^H(\theta_1)\x}{|\as^H(\theta_1)\x|}$
\State counter $\gets$ counter+1.
\EndWhile 
\end{algorithmic}
\end{algorithm}


The proposed algorithm gives the precise optimal solution of (\ref{eq:CLASSO}) which as we show by simulations is a proper robust approximation of the ML estimator. However, the searching steps might still be costly for certain fast applications. In this case, we introduce a different method of solving (\ref{eq:CLASSO}) which we call C-LASSO$_h$. 

The idea in this modification is that the long jumps in the LASSO path by the attractor $F$ could be substituted by a smoother development of $\lambda$. Although it may take more steps to achieve the next singular point, the simplicity of computations at each step compensates the complicated process at searching steps of C-LASSO. Note that the conditions in (\ref{eq:CLASSO}) imply that the spectrum $f(\theta)=|\as^H(\theta)\nhat|$ is bounded by $\lambda$ and achieves this value at the DOA estimates. As a graphical view, assume the graph $y=f(\theta)$ and the line $y=\lambda$ as the limit line touching the graph from above at DOA estimates. Decreasing $\lambda$ gradually, the limit line decreases and the graph gets more compressed. Roughly speaking, some points in the graph resist decreasing forming a peak which eventually touches the limit line. At this RP value a new DOA at the new touching point is introduced, which later follows decreasing with the limiting line. Thus, we can check our distance to the next singular point when decreasing $\lambda$ by looking at the highest peak point different to the estimates. If this peak is smaller than $\lambda$ we can continue our pass with a local change of parameters. However, at the next singular point the new touching point should be added to the active solutions. Otherwise, the extra peak will be more than $\lambda$ which indicates a wrong solution and a need to increase $\lambda$. Accordingly, we first introduce the following function:
\subsubsection{Local Equilevel Attractor (H)}
Given a quadruple $(I,r,\alpha,\lambda)$, the function $H$ gives the convergence point $(I^\prime,r^\prime,\alpha^\prime)$ of the Newton iterative solution of the system of equations given by the second line of (\ref{eq:CLASSO}) and (\ref{eq:mid}), i.e.
\begin{eqnarray}
&\as^H(\theta^\prime_i)\nhat^\prime=\lambda\frac{{s^\prime_i}}{|{s^\prime_i}|}\nonumber\\
&\Re(e^{-j\alpha^\prime_i}\ds^H(\theta^\prime_i)\nhat^\prime)=0.
\end{eqnarray}
The function $H$ can be computed fast using the Newton algorithm similar to the ML solution in Algorithm \ref{alg:ML} substituting 
$\J_\text{ML}$ and $\betta_\text{ML}$ by $\J_\text{LEA}$ and $\betta_\text{LEA}$ in (\ref{eq:JLEA}) and (\ref{eq:ettaLEA}) respectively. The C-LASSO$_h$ algorithm can then be expressed as in Algorithm \ref{alg:C-LASSOh}, where $\mu$ is a parameter setting the compromise between the speed and the probability of convergence. 

\begin{algorithm}
\caption{C-LASSO$_h$}
\label{alg:C-LASSOh}
\begin{algorithmic}
\State $\theta_0\gets\argmax\limits_{\theta}|\as^H(\theta)\x|$
\State $I\gets\{\theta_0\}$ and $\lambda\gets|\as^H(\theta_0)\x|$ 
\State $r\gets 0$ and $e^{j\alpha}\gets\frac{\as^H(\theta_0)\x}{|\as^H(\theta_0)\x|}$
\State counter $\gets$ 0.
\While {$\text{counter} < n$}
\State $\nhat\gets\x-\A(I)\s$
\State $\theta_1\gets\argmax\limits_\theta|\as^H(\theta)\nhat|$ s.t. $\theta\notin I$ is a local maximum. 
\State $p\gets|\as^H(\theta_1)\nhat|$
\If{$p=\lambda$}
\State $I\gets I\cup{\theta_1}$ 
\State $r_{\theta_1}\gets 0$ and $e^{j\alpha(\theta_1)}\gets\frac{\as^H(\theta_1)\x}{|\as^H(\theta_1)\x|}$
\State counter $\gets$ counter+1.
\Else
\State $\lambda\gets\mu\lambda+(1-\mu)p$
\State $(I, r,\alpha)\gets H(I,r,\alpha,\lambda)$
\EndIf
\EndWhile 
\end{algorithmic}
\end{algorithm}

%
\section{Numerical Results}
\begin{figure}[t]
\centering
\includegraphics[width=90mm]{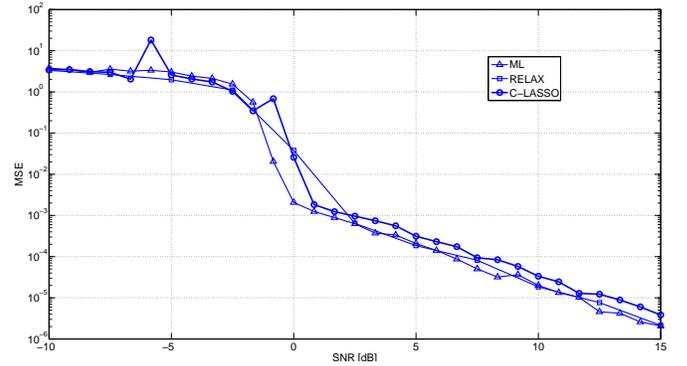}
\caption{Comparing MSE Vs SNR for 100 trials of single-snapshot data generated from a half wavelength ULA of m=15 sensors and 2 sources separated by the electrical angle $\frac{4\pi}{m}$.}
\label{fig:MSE}
\end{figure}

\begin{figure}[t]
\includegraphics[width=90mm]{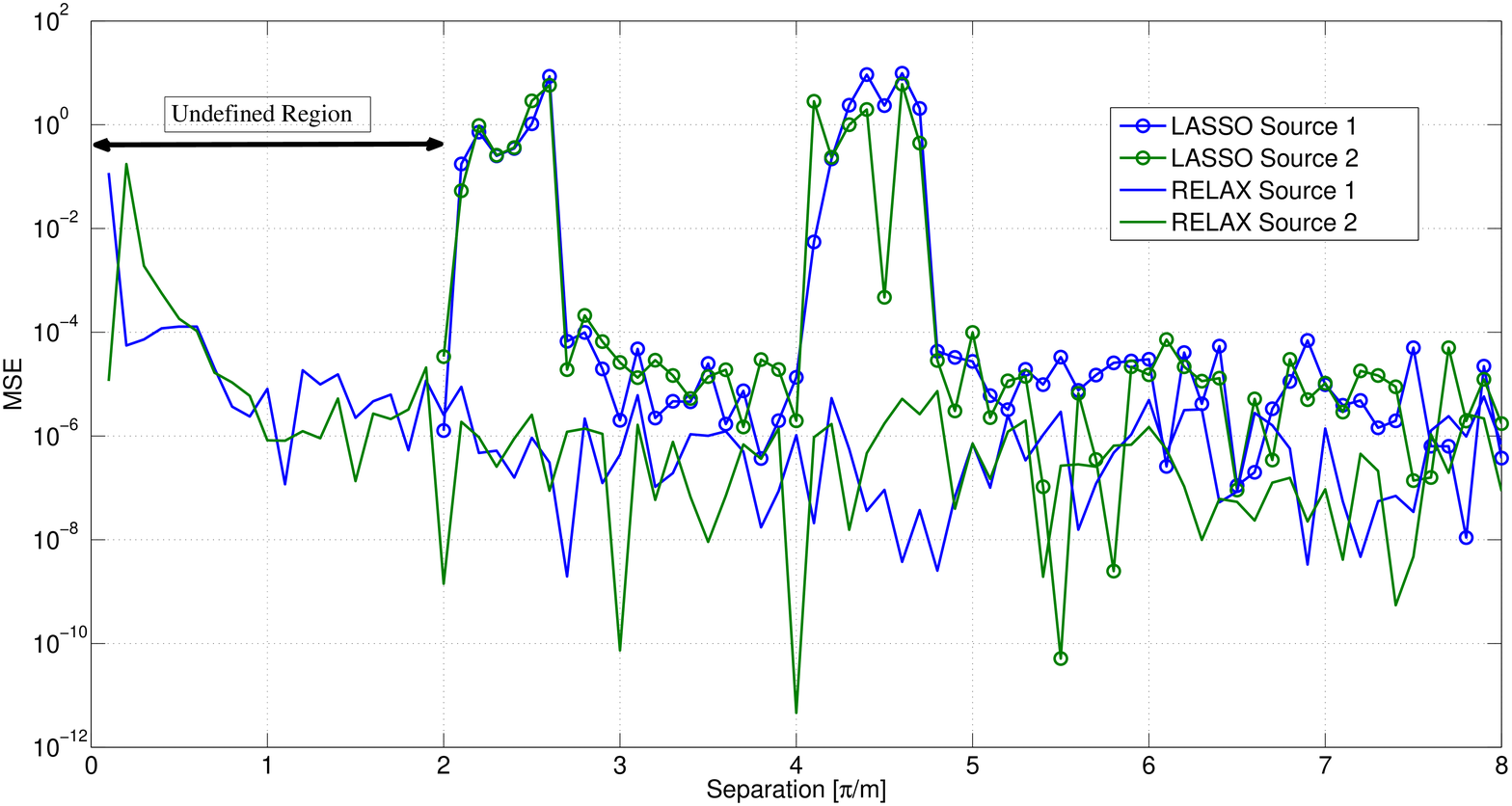}
\caption{Comparing MSE Vs separation in electrical angles for 100 trials of single-snapshot data generated from a half wavelength ULA of m=15 sensors. SNR=10 dB.}
\label{fig:NLLS}
\centering
\end{figure}

Assuming a standard Uniform Linear Array , we compared the result of applying C-LASSO to the single snapshot DOA model with the ML and RELAX estimator. The RELAX method alternates between estimating different subsets of DOAs using the previous estimate of the complement subset. In this work we use a singleton subset at each iteration to estimate which is often the case in practice. The ML estimator is implemented by first an exhaustive search and then the local Newton method solving (\ref{eq:KKTml}). Figure \ref{fig:MSE} shows the estimation Mean Squared Error (MSE) for these three techniques in the case of two sources with $s_1=s_2=1$ and separation $\Delta\phi=\frac{4\pi}{m}$, when the number of sensors $m$ is 15. As expected, LASSO has an estimation error level higher than MSE due to its bias. The RELAX estimate gives closer solution in this case. However, the C-LASSO reaches the threshold region at slightly lower SNR compared to RELAX.
Figure \ref{fig:NLLS} shows the error result of LASSO compared to RELAX fixing SNR to 10 dB and varying the separation in terms of electrical angle in the previous scenario. It shows a typical behavior of LASSO in the high SNR case, in which very close sources are absorbed to one source so that the LASSO solution path does not contain any point with correct model order. This is shown as the "undefined region" in Figure \ref{fig:NLLS}. However, the LASSO solution shows instabilities until separation reaches the fundamental resolution \cite{panahi_res}. 

\begin{figure}[t]
\centering
\includegraphics[width=90mm, height=45mm]{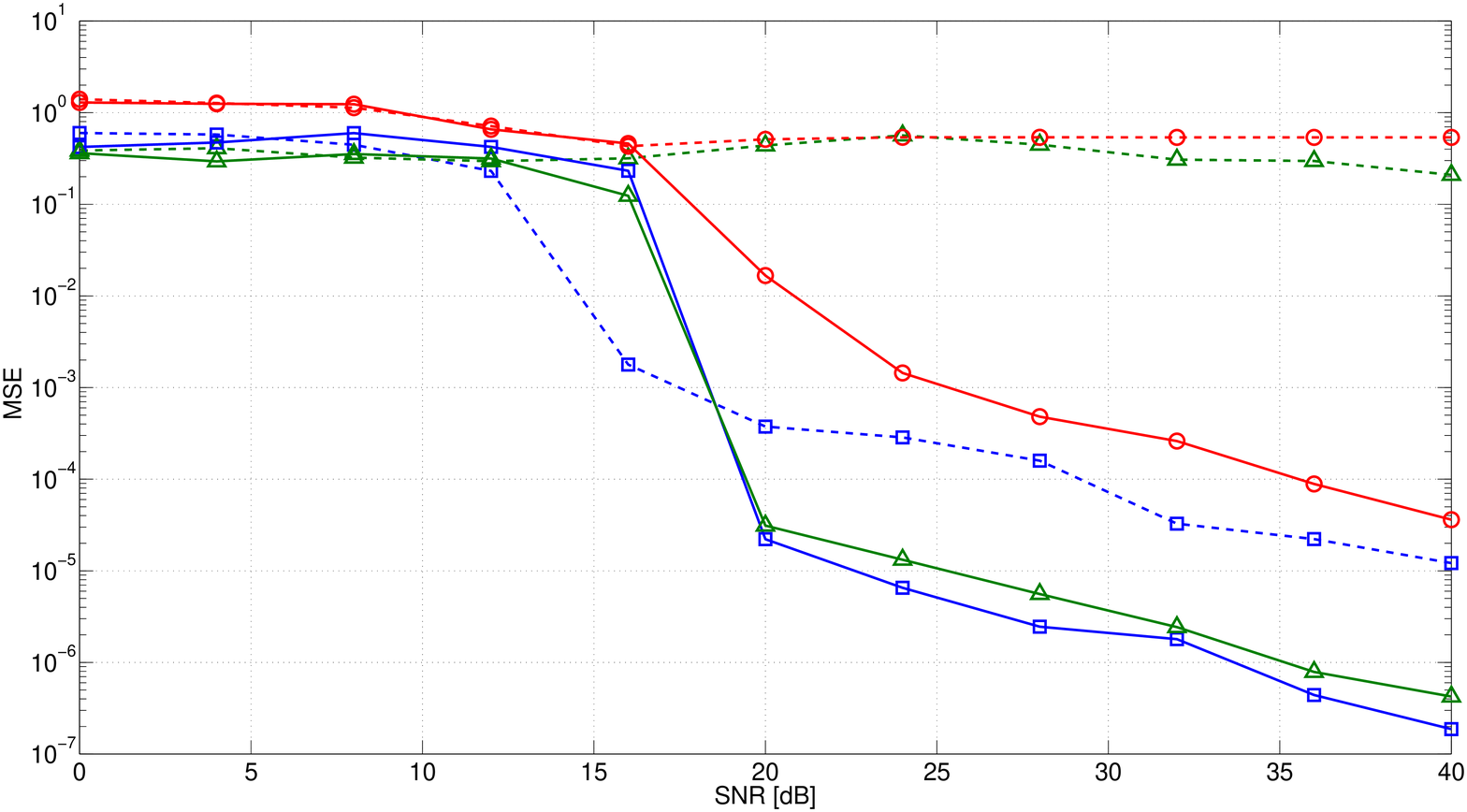}
\caption{C-LASSO performance compared to RELAX. The square, triangle, and circle markers show the DOA estimates of the first, second and third source respectively while the dashed lines show the result for RELAX.}
\label{fig:RELAX}
\end{figure}

In another experiment, we compared the C-LASSO$_h$ and RELAX performance in a more complicated case of three close sources at electrical angles $[\frac{-5\pi}{m}\ 0\  \frac{3.5\pi}{m}]$ with $s_1=s_2=1$ and $s_3=0.1j$. The SNR is measured by observing $s_1$ and $n_1$ when $\mu=0.8$. The scenario is hard for the RELAX algorithm in the one snapshot case due to the  different levels of sources. The results are shown in Figure \ref{fig:RELAX}. As can be seen, the RELAX technique can only resolve the reference signal $s_1$, while too wide dynamic range makes it hard to estimate the DOAs of the second and the third sources respectively. However, one may note the higher threshold SNR of C-LASSO$_h$. Note also that the realization of C-LASSO$_h$ took 60 times longer time than RELAX using MATLAB programming. 

\section{Conclusion}
 In this work, we introduced a LASSO realization technique, C-LASSO, in the one snapshot case and an improvement C-LASSO$_h$ from the complexity point of view which do not need any discretization. They are much faster than the convex programming realizations, such as the interior point technique considering the problem of RP selection. These algorithms are based on the idea of following the optimality point path of a set of generalized optimality conditions derived from the LASSO original ones.

The results show that in lower SNRs and more sources case, C-LASSO dominates ML from complexity point of view without loosing much performance. Although there exists other DOA estimation techniques approximating the ML solution, we showed by simulation that the LASSO algorithm is more robust to the problem of wide dynamic range .  It can be shown, while neglected in this work, that the LASSO bias is linear with the noise level independent of the true DOAs and sources.

The C-LASSO algorithms drawback is its computational time due to one dimensional search steps and the parameter $\mu$ in C-LASSO$_h$, which could not be decreased arbitrarily due to the convergence problem. However, sacrificing some performance, one may confine the search to a sufficiently fine grid neglecting the fine tuning step which can speed up the algorithm. Furthermore, while we have not presented the mathematical details here, LASSO encounters a consistency problem in the one snapshot case of the sources separated less than a fundamental resolution (see \cite{panahi_res} for a similar argument).

Finally, it should be noted that similar to the Group-LASSO (G-LASSO) formalism, C-LASSO could be adapted to multiple-snapshot model. It is expected that due to the good performance of LASSO in high SNRs, the grouped C-LASSO also provides a robust technique.
\bibliographystyle{IEEEtran}
\bibliography{bibFile}

\begin{thebibliography}{10}
\providecommand{\url}[1]{#1}
\csname url@samestyle\endcsname
\providecommand{\newblock}{\relax}
\providecommand{\bibinfo}[2]{#2}
\providecommand{\BIBentrySTDinterwordspacing}{\spaceskip=0pt\relax}
\providecommand{\BIBentryALTinterwordstretchfactor}{4}
\providecommand{\BIBentryALTinterwordspacing}{\spaceskip=\fontdimen2\font plus
\BIBentryALTinterwordstretchfactor\fontdimen3\font minus
  \fontdimen4\font\relax}
\providecommand{\BIBforeignlanguage}[2]{{%
\expandafter\ifx\csname l@#1\endcsname\relax
\typeout{** WARNING: IEEEtran.bst: No hyphenation pattern has been}%
\typeout{** loaded for the language `#1'. Using the pattern for}%
\typeout{** the default language instead.}%
\else
\language=\csname l@#1\endcsname
\fi
#2}}
\providecommand{\BIBdecl}{\relax}
\BIBdecl

\bibitem{tibshirani}
R.~Tibshirani, ``Regression shrinkage and selection via the lasso,''
  \emph{Journal of the Royal Statistical Society, Series B, (Methodological)},
  vol.~58, no.~1, pp. 267--288, 1996.

\bibitem{Donoho_CS}
D.~Donoho, ``Compressed sensing,'' \emph{Information Theory, IEEE Transactions
  on}, vol.~52, no.~4, pp. 1289 --1306, april 2006.

\bibitem{Malioutov}
D.~Malioutov, M.~Cetin, and A.~Willsky, ``Source localization by enforcing
  sparsity through a laplacian prior: an svd-based approach,'' \emph{IEEE
  Workshop on Statistical Signal Processing}, pp. 573 -- 576, sept.-1 oct.
  2003.

\bibitem{Panahi1}
A.~Panahi and M.~Viberg, ``Maximum aposteriory based regularization parameter
  selection,'' in \emph{International Conference on Acoustics, Speech, and
  Signal Processing}, 2011.

\bibitem{BLASSO}
T.~Park and G.~Casella, ``The bayesian lasso,'' \emph{Journal of the American
  Statistical Association}, vol. 103, p. 681 – 686, 2008.

\bibitem{cvx}
M.~Grant and S.~Boyd, ``{CVX}: Matlab software for disciplined convex
  programming, version 1.21,'' \url{http://cvxr.com/cvx}, Apr. 2011.

\bibitem{LARS}
B.~Efron, T.~Hastie, L.~Johnstone, and R.~Tibshirani, ``{Least angle
  regression},'' \emph{Annals of Statistics}, vol.~32, pp. 407--499, 2004.

\bibitem{Homotopy}
M.~R. Osborne, B.~Presnell, and B.~Turlach, ``A new approach to variable
  selection in least squares problems,'' 1999.

\bibitem{panahi_fast}
A.~Panahi and M.~Viberg, ``Fast candidate points selection in the lasso path,''
  \emph{Signal Processing Letters, IEEE}, vol.~19, no.~2, pp. 79 --82, feb.
  2012.

\bibitem{SAGE}
J.~Fessler and A.~Hero, ``Space-alternating generalized
  expectation-maximization algorithm,'' \emph{Signal Processing, IEEE
  Transactions on}, vol.~42, no.~10, pp. 2664 --2677, oct 1994.

\bibitem{Viberg_2decades}
H.~Krim and M.~Viberg, ``Two decades of array signal processing research: the
  parametric approach,'' \emph{Signal Processing Magazine, IEEE}, vol.~13,
  no.~4, pp. 67 --94, jul 1996.

\bibitem{Bazaraa}
M.~S. Bazaraa, H.~D. Sherali, and C.~M. Shetty, \emph{Nonlinear Programming:
  Theory and Algorithms; 3rd ed.}\hskip 1em plus 0.5em minus 0.4em\relax
  Newark, NJ: Wiley, 2006.

\bibitem{Newtonian}
D.~Starer and A.~Nehorai, ``Newton algorithms for conditional and unconditional
  maximum likelihood estimation of the parameters of exponential signals in
  noise,'' \emph{Signal Processing, IEEE Transactions on}, vol.~40, no.~6, pp.
  1528 --1534, jun 1992.

\bibitem{RELAX}
\BIBentryALTinterwordspacing
Z.-S. Liu, J.~Li, and P.~Stoica, ``Relax-based estimation of damped sinusoidal
  signal parameters,'' \emph{Signal Processing}, vol.~62, no.~3, pp. 311 --
  321, 1997. [Online]. Available:
  \url{http://www.sciencedirect.com/science/article/pii/S0165168497001321}
\BIBentrySTDinterwordspacing

\bibitem{panahi_res}
A.~Panahi and M.~Viberg, ``On the resolution of the lasso-based doa estimation
  method,'' in \emph{Smart Antennas (WSA), 2011 International ITG Workshop on},
  feb. 2011, pp. 1 --5.

\end{thebibliography}
%
%
%

\end{document}